\documentstyle[sprocl]{article} 
 
\def\be{\begin{equation}} 
\def\ee{\end{equation}} 
\def\bea{\begin{eqnarray}} 
\def\eea{\end{eqnarray}} 

\begin{document}


\def\hDots(#1,#2,#3){\multiput(#1,#2)(5,0){#3}{\circle*{3}}}
\def\hdots(#1,#2,#3){\multiput(#1,#2)(3,0){#3}{\circle*{1}}}
\def\Xdots(#1,#2){
   \multiput( #1, #1)( 2.12, 2.12){#2}{\circle*{1}}
   \multiput( #1,-#1)( 2.12,-2.12){#2}{\circle*{1}}
   \multiput(-#1, #1)(-2.12, 2.12){#2}{\circle*{1}}
   \multiput(-#1,-#1)(-2.12,-2.12){#2}{\circle*{1}}}
\def\bigcirc (#1,#2,#3){\thicklines\put(#1,#2){\circle{#3}}}
\def\fcircle (#1,#2,#3){\put(#1,#2){\circle*{#3}}}
\def\fselfen{\bigcirc (0,0,28)\hdots(-25,0,4)\hdots(16,0,4)}
\def\fvertex{\bigcirc (0,0,28)\Xdots(12,4)}
\def\Bbl#1#2{\multiput(#1 15,0)(#1 1.5,#2 2.5){7}{\circle*{1}}}
\def\Bloop{ \Bbl++  \Bbl+-  \Bbl-+  \Bbl-- 
\fcircle (-12, 4,3)\fcircle (-9, 8,3)\fcircle (-5, 11,3)\fcircle (0, 12,3)
\fcircle (-12,-4,3)\fcircle (-9,-8,3)\fcircle (-5,-11,3)\fcircle (0,-12,3)
\fcircle ( 12, 4,3)\fcircle ( 9, 8,3)\fcircle ( 5, 11,3)\fcircle ( 15,0,3)
\fcircle ( 12,-4,3)\fcircle ( 9,-8,3)\fcircle ( 5,-11,3)\fcircle (-15,0,3)  }
\def\rmchar #1{{\rm #1}}
\def\fsquare {\displaystyle}
\def\slash {\put(0.5,1){/}}
\hfill SMC-PHYS-152\vskip 3mm
\title{  COMPOSITENESS CONDITION IN THE NAMBU-JONA-LASINIO MODEL
\footnote{Invited talk presented at 1996 International Workshop
on Perspectives of Strong Coupling Gauge Theories, Nagoya, Japan, 
November 1996. To be published in Proceedings.}
}
\author{                   KEIICHI AKAMA
}
\address{  Department of Physics, Saitama Medical College,
\\             Kawakado, Moroyama, Saitama, 350-04, Japan
}  
\maketitle
\abstracts{ 
The Nambu-Jona-Lasinio model is the special case 
	of the renormalized Yukawa model 
	with the compositeness condition.
We use it to calculate the effective coupling constants 
	in terms of the compositeness sale (momentum cut off)
	at the next-to-leading order in $1/N$.
The next-to-leading correction is too large 
	in the model of scalar composite,
	while that in the induced gauge theory is reasonably suppressed
	due to the gauge cancellation of the leading divergences.
} 

\section {Introduction} 

The Nambu-Jona-Lasinio model \cite{NJL} realizes tractable compositeness 
	in a simple field-theoretical scheme. 
What is important there is the compositeness condition \cite{cc}
	which connect the Nambu-Jona-Lasinio model 
	to a special case of the renormalized Yukawa model.\cite{LM,ECCS}
However, the relations used there are mostly based on arguments
	which hold only in the large $N$ limit,
	where $N$ is the number of the fermion species.
In this talk, I would like to consider the compositeness condition
	$Z=0$ at the next-to-leading order in $1/N$ expansion,\cite{Akama,AH}
	where $Z$ is the wave-function renormalization constant
	of the `composite boson'.

About twenty years ago, Terazawa, Chikashige and myself \cite{TCA,SS}
	considered a Nambu-Jona-Lasinio type model of the standard model
	with composite Higgs scalar and gauge bosons 
	made of a quark-anti-quark pair.
It becomes a spontaneously broken gauge theory at low energies.
It provides various predictions 
 	on relations among the masses and coupling constants.
According to them, at least one of the quarks 
 	should have a mass of the order of the weak interaction scale.
It looked puzzling 
 	because the known quarks at that time 
 	were much smaller than the weak scale.
Today, however, we know that 
 	the top quark has the mass of the order of the weak scale,\cite{top}
 	and the sum rule becomes rather natural.
This fact called the revived attentions to the NJL-type model 
	of the spontaneously broken electroweak symmetry.\cite{MTY,BHL}
Numerically, however, it does not precisely hold.
We need to consider how to make it more precise
	beyond the leading approximation in ${1/N}$. \cite{precise}

\section {Compositeness Condition in the NJL model} 
We consider the NJL model for the fermion 
$\psi =\{\psi _1,\psi _2,\cdots ,\psi _N\}$
	with $N$ colors given by the Lagrangian 
\begin{eqnarray} 
  {\cal L}_\rmchar {NJL}=\overline \psi i\slash \partial \psi  
   + f|\overline \psi _\rmchar L\psi _\rmchar R|^2
\end{eqnarray} 
	with $U(1)\times U(1)$ chiral symmetry,
	where $f$ is the coupling constant.
In 3+1 dimensions, it is not renormalizable,
	and we assume a very large but finite momentum cutoff.
This Lagrangian ${\cal L}_\rmchar {NJL}$ is known to be 
equivalent to \cite{GN,KK}
\begin{eqnarray} 
{\cal L}'_\rmchar {NJL}=\overline \psi i\slash \partial \psi 
	  +(\overline \psi _\rmchar L \phi  \psi _\rmchar R+\rmchar {h.c.})
	  -{1\over f}|\phi |^2
\end{eqnarray} 
written in terms of the auxiliary boson field $\phi $.
Now compare it with this Lagrangian of the renormalized Yukawa model
\begin{eqnarray} 
{\cal L}_\rmchar {Yukawa}&=&
	Z_\psi \overline \psi _\rmchar ri\slash \partial \psi _\rmchar r
	+Z_g g_\rmchar r
	(\overline\psi_\rmchar{rL}\phi_\rmchar r\psi_\rmchar{rR}
	+\rmchar {h.c.})\cr 
&&        
	+Z_\phi |\partial _\mu \phi _\rmchar r|^2  
	-Z_\mu   \mu _\rmchar r^2|\phi _\rmchar r|^2 
	-Z_\lambda \lambda _\rmchar r|\phi _\rmchar r|^4,
\end{eqnarray} 
where $\psi _\rmchar r$ and $\phi _\rmchar r$ are 
	the renormalized fermion and boson fields, respectively, 
	$g_\rmchar r$ and $\lambda _\rmchar r$ are 
	the renormalized coupling constants, 
	$\mu _\rmchar r$ is the renormalized mass 
	of the field $\phi _\rmchar r$,
	and $Z_\psi $, $Z_\phi $, $Z_g$, $Z_\mu $, and $Z_\lambda $
	are the renormalization constants.
We can see that, if 
\begin{eqnarray} 
	Z_\phi =Z_\lambda =0		\label{cc},
\end{eqnarray} 
	the Lagrangian ${\cal L}_\rmchar {Yukawa}$ 
	coincides with ${\cal L}'_\rmchar {NJL}$, \cite{ECCS}
	where we identify $\psi $, $\phi $, 
	and $f$ in ${\cal L}'_\rmchar {NJL}$,
	with $\sqrt {Z_\psi }\psi _\rmchar r$, 
	$(Z_g /Z_\psi )g_\rmchar r\phi _\rmchar r$, and
	$Z_g^2g^2_\rmchar r/Z_\psi ^2Z_\mu  \mu _\rmchar r^2$ 
	in ${\cal L}_\rmchar {Yukawa}$, respectively.
The condition (\ref{cc}) is called `compositeness condition'.\cite{cc}
Thus the Lagrangian for NJL model is the special case 
	of the renormalized Yukawa model with the compositeness condition.
The compositeness condition gives rise to relations
	among coupling constants $g_\rmchar r$, $\lambda _\rmchar r$, 
and the cut off $\Lambda $.
If the chiral symmetry is spontaneously broken,
	they imply relations among the fermion mass $m_f$, 
	the Higgs-scalar mass $M_H$, and the cutoff $\Lambda $.
Thus we can study everything in the NJL model
	by studying the well-understood Yukawa model,
	and by imposing the compositeness condition 
	on the coupling constants and masses. 
Then what is urgent is 
	to work out the compositeness condition,
	and solve it for the coupling constants.

\subsection {Lowest order in $1/N$} 
For an illustration, we begin with 
	the lowest-order contributions in $1/N$ expansion.
In the Yukawa model, the boson self-energy part 
	and the four-boson vertex part are given by the diagrams
\vskip 15pt

\begin{picture}(80,0)       
\thicklines
\put(0,0){                  
  \put( 50,3){  \fselfen }
  \put( 85,0){  $=g_\rmchar r^2NIp^2$, }
  \put(175,3){  \hdots(-18,0,13) 
                  \put(-5,-5){\line(1, 1){10}}
                  \put(-5, 5){\line(1,-1){10}}  }
  \put(200,0){  $=(Z_\phi -1)p^2$,}
}
\put(0,-40){                
  \put( 50,3){  \fvertex }
  \put( 85,0){  $=g_\rmchar r^4NI$,}
  \put(175,3){  \Xdots(0,7)
                  \put(0,-7){\line(0,1){14}}
                  \put(-7,0){\line(1,0){14}}  }
  \put(200,0){  $=(Z_\lambda -1)\lambda _\rmchar r$,}
}
\end{picture}
\vskip 55pt\noindent 
where the solid (dotted) line indicates 
	the fermion (boson) propagator, 
and
\begin{eqnarray} 
I=\cases{
{\fsquare {1\over 16\pi ^2}{1\over \epsilon }}\ \  
	\rmchar {(dimensional\ regularization,}\ 
	\epsilon =\fsquare {4-d\over 2},\ d:{\rm dimension})
\cr 
\fsquare {1\over 16\pi ^2}\log\Lambda ^2\ \  
   ({\rm Pauli\ Villars\ regularization,}\ \Lambda :{\rm regulator\ mass})
}
\end{eqnarray} 
The renormalization constants $Z_\phi $ and $Z_\lambda $ 
	should be chosen as
\begin{eqnarray} 
Z_\phi =1-g_\rmchar r^2NI, \ \ \ \ 
Z_\lambda \lambda _\rmchar r=\lambda _\rmchar r - g_\rmchar r^4NI,
\end{eqnarray} 
so as to cancel out all the divergences.
Then the compositeness condition is obtained 
	by putting $Z_\phi =Z_\lambda =0$,
	and it is easily solved to give \cite{ECCS}
\begin{eqnarray} 
    g_\rmchar r^2={1\over NI}, \ \ \ \ \lambda _\rmchar r={1\over NI}
\end{eqnarray} 
If the chiral symmetry is spontaneously broken 
	(i.e. if $\mu _\rmchar r^2<0$),
	the physical fermion mass $m_f$ and 
	the physical Higgs mass $M_H$ are given by
\begin{eqnarray} 
    m_f=g_\rmchar r\langle \phi \rangle 
	=\langle \phi \rangle /\sqrt {NI}, \ \ \ 
    M_H=2\sqrt \lambda _\rmchar r\langle \phi \rangle 
	=2\langle \phi \rangle /\sqrt {NI}, \ \ \ 
\end{eqnarray} 
and hence we have $2m_f=M_H$.
These reproduce the well known results of the lowest order 
	Nambu-Jona-Lasinio model.\cite{NJL}

\subsection {Next-to-leading order in $1/N$} 
Now we turn to the next-to-leading order in $1/N$.
\footnote{This section reviews Ref.\put(4,-4.5){\large\cite{Akama}}}
In the Yukawa model, 
	the boson self-energy part is given by the diagram
\vskip 12pt
\begin{picture}(50,0)      
\put(20,3){ \fselfen\fcircle (-9,6,3)
    \fcircle (-5,3,3)\fcircle (0,2,3)\fcircle (5,3,3)\fcircle (9,6,3) }
\end{picture}
+ the counter terms for all the subdiagram divergences,
\vskip 12pt\noindent 
where
\begin{picture}(220,0)     
\hDots(5,3,5)
\put(35,0){stands for \hdots(5,3,9)}
\put(114,0){+}
\put(125,0){\hdots(2,3,3)\bigcirc (15,3,10)\hdots(22,3,3)}
\put(160,0){+}
\put(171,0){\hdots(2,3,3)\bigcirc (15,3,10)\hdots(22,3,3)
			 \bigcirc (35,3,10)\hdots(42,3,3)}
\end{picture}
+ $\cdots $.
The renormalization constant $Z_\phi $ is calculated to be 
\begin{eqnarray} 
Z_\phi =1-g_\rmchar r^2NI-g_\rmchar r^2I
	-{1\over N}(1-g_\rmchar r^2NI)\log(1-g_\rmchar r^2NI)
\end{eqnarray} 
	so as to cancel out all the divergences there. 
The logarithm arises from the infinite sum 
	over the fermion loop insertions into the internal boson line.
Similarly, the four boson vertex part is given by the diagrams
\vskip 15pt
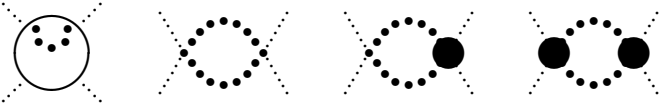
\begin{picture}(80,0)     
\put( 20,3){  \fvertex\fcircle (-6,9,3) \fcircle (-5,4,3) 
	\fcircle (0,2,3) \fcircle (5,4,3) \fcircle (6,9,3) }
\put( 85,3){  \Bloop} 
\put(155,3){  \Bloop\fcircle (15,0,12)}
\put(225,3){  \Bloop\fcircle (-15,0,12) \fcircle (15,0,12)}
\end{picture}
\vskip 15pt\noindent 
and the counter terms for all the subdiagram divergences,
where \fcircle (10,3,12)\hskip 20pt stands for 
\\\bigcirc (10,3,12)\hskip 20pt
with arbitrary permutations of four external boson lines. 
The renormalization constant $Z_\lambda $ is calculated to be 
\begin{eqnarray} &
Z_\lambda \lambda _\rmchar r&=\lambda _\rmchar r 
	- g_\rmchar r^4NI+8g_\rmchar r^4I
	+{20(\lambda _\rmchar r - g_\rmchar r^2)^2I\over 1-g_\rmchar r^2NI}
\cr && 
	-{1\over N}\left[ 2g_\rmchar r^2(1-g_\rmchar r^2NI)
	+20(\lambda _\rmchar r - g_\rmchar r^2)\right] 
	\log(1-g_\rmchar r^2NI)
\end{eqnarray} 
	so as to cancel out all the divergences here, 
	where the logarithm again arises from the infinite sum 
	over the fermion loop insertions into the internal boson lines.
The compositeness condition is given 
	by putting these expressions vanishing.
Though it looks somewhat complex at first sight,
	it can be solved by iteration to give the very simple solution
\begin{eqnarray} 
g_\rmchar r^2={1\over NI}\left[ 1-{1\over N}+O({1\over N^2})\right] ,
\ \ \ \ \
\lambda _\rmchar r ={1\over NI}\left[ 1-{10\over N}+O({1\over N^2})\right] .
\end{eqnarray}

If the chiral symmetry is spontaneously broken
	the masses of the physical fermion and physical Higgs scalar
	are given by 
\begin{eqnarray} 
m_f&=&g_\rmchar r\langle \phi \rangle 
	={\langle \phi \rangle \over \sqrt {NI}}
	\left[ 1-{1\over 2N}+O({1\over N^2})\right] ,\cr  
M_H&=&2\sqrt {\lambda _\rmchar r}\langle \phi \rangle 
	={2\langle \phi \rangle \over \sqrt {NI}}
	\left[ 1-{5\over N}+O({1\over N^2})\right] ,\cr   
{\rm hence}&& {M_H\over m_f}=2\left[ 1-{9\over 2N}+O({1\over N^2})\right] .
\end{eqnarray} 
For the case of $N=3$ of the practical interest,
	the corrections turn out to be too large,
	and the coupling constant $\lambda $ is negative,
	which implies that the Higgs potential is unstable.
The origin of this large negative contributions 
	is traced back to the boson loop diagrams.
A possible way getting rid of these difficulties
	is to assume that the cutoff $\Lambda _\phi $ 
	for the composite $\phi $
	is much smaller than the cutoff $\Lambda $ 
	for the elementary fermions.
In this case, the correction terms are suppressed by the small factor 
	$r=\log\Lambda _\phi /\log\Lambda $. 
\begin{eqnarray} 
g_\rmchar r^2={1\over NI}
	\left[ 1-{r\over N}+O({r^2\over N^2})\right] ,
\ \ \ \ 
\lambda _\rmchar r ={1\over NI}
	\left[ 1-{10r\over N}+O({r^2\over N^2})\right] .
\end{eqnarray}

It is straightforward to extend these results 
	to the models of larger chiral symmetries.
For example, for $SU(2)\times SU(2)$ for the pions and $\sigma $ meson,
	the correction to $g_\rmchar r$ is absent at this order,
	and that for $\lambda _\rmchar r$ is $6/N$.
For $SU(2)\times U(1)$ of the electroweak symmetry,
	they are $3/2N$ and $12/N$, respectively.
The corrections for the cases of the general number $F$ of flavors 
	are also calculated.
In any case, the corrections are too large for $N=3$.

\section {Compositeness Condition in the induced gauge theory} 
We can apply this method to the induced gauge theory,
	namely, the gauge theory with a composite gauge field.
\footnote{This section reviews Ref.\put(4,-4.5){\large\cite{AH}}}
It is given by the strong coupling limit $f\rightarrow \infty $ of the
	vector-type four Fermi interaction model \cite{HBB}
\begin{eqnarray} 
   {\cal L}_\rmchar {4F}
	=\overline \psi _j \left( i\slash \partial -m\right) \psi  
	-f\left( \overline \psi \gamma _\mu \psi \right) ^2,
\end{eqnarray} 
where $\psi =\{\psi _1,\psi _2,\cdots ,\psi _N\}$, $f$ 
	is the coupling constant,
	and $m$ is the mass of $\psi $.
The Lagrangian is equivalent to 
	the linearized one written in terms 
	of the auxiliary vector-boson field $A_\mu $.
We can see that it is the special case of the renormalized gauge theory
	with the compositeness condition $Z_3=0$,
	where $Z_3$ is the wave-function renormalization constant
	of the gauge field identified with $A_\mu$.
In the gauge theory, 
	$Z_3$ should be chosen so as to cancel out the divergences 
	in the gauge boson self-energy part.
At the leading and next-to-leading order, 
	it is given by the following diagrams.
\vskip 25pt
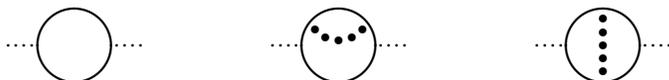
\begin{picture}(0,0)       
\put(50,15){ \fselfen }
\put(150,15){ \fselfen \fcircle (-9,6,3)\fcircle (-5,3,3)\fcircle (0,2,3)
			\fcircle (5,3,3)\fcircle (9,6,3)}
\put(250,15){ \fselfen \fcircle (0,-10,3)\fcircle (0,-5,3)\fcircle (0,0,3)
			\fcircle (0,5,3)\fcircle (0,10,3)}
\end{picture}
\\
After a lengthy calculation, we obtain 
\begin{eqnarray} 
Z_3=1-{e_{\rm r}^2 N \over 12\pi ^2 \epsilon  } 
    -{3e_{\rm r}^2 \over 16\pi ^2 }
	\left[  1+\left(  1-{12\pi ^2\epsilon \over e_{\rm r}^2 N }\right)  
	\ln\left(  1-{e_{\rm r}^2 N \over 12\pi ^2\epsilon }\right) \right] ,
\end{eqnarray} 
where $e_{\rm r}$ is the renormalized effective coupling constant.
Then, the compositeness condition $Z_3=0$ 
	is solved to give the simple solution
\begin{eqnarray} 
e_\rmchar r^2 = { 12 \pi ^2 \epsilon \over N }
	\left[  1-{9\epsilon \over 4 N }+O({1\over N^2})\right] .
\end{eqnarray} 
The correction term $9\epsilon /4N$ 
	is naturally suppressed by the small factor $\epsilon $.
It justifies the lowest order approximation of this model
	unlike in the case of the 
	aforementioned NJL model of the scalar composite.
The origin of the suppression factor is traced back to 
	the gauge cancellation of the leading divergence 
	in the next-to-leading order diagrams.
So far we assumed that all the fermions have the same charges for simplicity.
If the charges $Q_j$ are different, the expression is modified as follows.
\begin{eqnarray} 
e_\rmchar r^2 ={ 12 \pi ^2 \epsilon \over \sum _j Q_j^2 }
	\left[  1-{9\epsilon \sum _j Q_j^4\over 4(\sum _j Q_j^2)^2}\right] .
\end{eqnarray} 
If we apply this to the quantum electrodynamics
	with 3 generations of quarks and leptons,
	$\epsilon $ is estimated to be 6$\times 10^{-3}$,
	which implies the next-to-leading order correction amounts 
	only to 0.1\% of the lowest order term.

\section {Summary} 
The Nambu-Jona-Lasinio model is the special case 
	of the renormalized Yukawa model 
	with the compositeness condition $Z_\phi =Z_\lambda =0$.
We used it to calculate the effective coupling constants 
	in terms of the compositeness sale (momentum cut off)
	at the next-to-leading order in $1/N$.
The next-to-leading correction to 
	the coupling constant $g_\rmchar r^2$ is $1/N$,
	and that to the coupling constant $\lambda _\rmchar r$ is $10/N$.
For $N=3$ of our practical interests,
	the corrections are too large,
	and $\lambda <0$, as implies unstable Higgs potential.
For induced gauge theory, 
	the compositeness condition $Z_3=0$ implies that
	the next-to-leading correction term is $9\epsilon /4N$, 
	which is naturally suppressed by the small factor $\epsilon $.
Interesting extensions to the nonabelian gauge theories 
	are now under investigation.\cite{AH2}  
In this case,
	if the corresponding elementary gauge theory 
	is asymptotically free,
	the next-to-leading corrections 
	according to the compositeness condition
	become too large to justify the $1/N$ expansion.
Finally we comment that,
	the compositeness condition holds independently 
	of choice of the renormalization conditions, 
	because the renormalization is multiplicative.
It implies the NJL model is at the fixed point
	in the renormalization flow of the Yukawa model
	(with a fixed cutoff).
This is consistent with the fact that the compositeness condition
	is a relation among the observable renormalized quantities.

\section *{Acknowledgments} 
I would like to thank Professor H.~Terazawa for stimulating discussions,
Professor T.~Hattori for collaborations,
and Professor K.~Yamawaki and the organizing committee of the workshop 
for arranging the opportunity for me to present this work.

\section *{References} 

 \end{document}